\documentclass[preprintnumbers,amsmath,amssymb,floatfix,10pt,prd,onecolumn,
superscriptaddress,nofootinbib]{revtex4}
\usepackage{latexsym}
\usepackage{epsfig}
\usepackage{epstopdf}
\usepackage{graphicx}
\usepackage{amssymb}
\usepackage{amsmath}
\usepackage{dcolumn}
\usepackage{bm}
\usepackage{color}
\usepackage{comment}
\usepackage[utf8]{inputenc}
\begin{document}

\title{\bf An anisotropic version of Tolman VII solution in $f(R,T)$ gravity via Gravitational Decoupling MGD Approach}

\author{Hina Azmat}
\email{hinaazmat0959@gmail.com}\affiliation{Department of Mathematics, COMSATS University Islamabad, Lahore Campus, Lahore, Pakistan}

\author{M. Zubair}
\email{mzubairkk@gmail.com;drmzubair@cuilahore.edu.pk}\affiliation{Department of Mathematics, COMSATS University Islamabad, Lahore Campus, Lahore, Pakistan}

\begin{abstract}
In this work, we have adopted gravitational decoupling by Minimal Geometric Deformation (MGD) approach and have developed an anisotropic version of
well-known Tolman VII isotropic solution in the framework of $f(R, T)$ gravity, where $R$ is
Ricci scalar and $T$ is trace of energy momentum tensor.
The set of field equations has been developed with respect to total energy momentum tensor, which combines
effective energy momentum tensor in $f(R,T)$ gravity and additional source $\phi_{ij}$.
Following MGD approach, the set of field equations has been separated into two sections. One
section represents $f(R, T)$ field equations, while the other is related to the source $\phi_{ij}$.
The matching conditions for inner and outer geometry have also been discussed and an anisotropic solution
has been developed using mimic constraint for radial pressure.
In order to check viability of the solution, we have considered observation data of three different
compact star models, named PSR J1614-2230, PSR 1937+21 and SAX J1808.4-3658 and have discussed thermodynamical properties analytically and graphically.
 The energy conditions are found to be satisfied for the three compact stars.
The stability analysis has been presented through causality condition and Herrera's cracking concept,
which ensures physical acceptability of the solution.
\end{abstract}

\maketitle

\date{\today}

\section{Introduction}

The expansion of universe which has widely been accepted through observational
data and the emergence of different structures in the universe
provide motivation for the study of cosmological evolution.
These phenomena are associated with the presence of dark components termed as dark matter and dark energy.
Dark matter is an important ingredient
for the dynamics of the entire universe which manifests its presence in
flattened galactic rotational curves, while dark energy is considered as a most strong candidate which is responsible
for the late time cosmic acceleration.
General Relativity (GR) is the most
comprehensive theory that is in excellent agreement with observational
data. However, there are still various issues in fundamental physics,
astrophysics and cosmology that indicate towards the shortcomings of GR and
advocate the necessity of
alternative approaches to the resolution of dark side puzzle.
These approaches are mainly based on two strategies, however modified
theories of gravity are considered most successful tool for the description of cosmic evolution.
These modified theories include Brans-Dicke, $f(R)$ theory, $f(G)$ theory, $f(R,T)$ theory and many others.

The $f(R,T)$ theory \cite{1} where matter lagrangian comprises on an arbitrary function of matter and
geometry was proposed by Harko and his collaborators, in which $R$ indicates Ricci scalar and $T$ represents the trace of energy momentum
tensor. The importance of $T$ in the theory can be observed by the exotic form of
matter or phenomenological aspects of quantum gravity. The equations of motion have been formulated by using the metric approach instead of the Palatini
approach. Due to the coupling between matter and geometry, the motion of test particles is found to be non-geodesic
and an additional acceleration is always there. The $f(R,T)$ theory meets weak-field solar system conditions successfully,
and new investigations make us able to assemble the galactic impacts of dark matter with the help of this theory \cite{2}. It has also been found
that $f(R,T)$ gravity shows clear tendency towards the deviation of the standard geodesic equation \cite{3}
and gravitational lensing \cite{4}. Due to its viable nature, it can be
applied to explore multiple astrophysical and cosmological phenomena, which can
give rise to significant developments.

Houndjo \cite{4*} worked out cosmological reconstruction of $f(R,T)$
gravity and examined the transition of matter dominated epoch to the late time accelerated
regime. Barrientos et al. \cite{5} studied the $f(R,T)$ theories of gravity
and its application with affine connection.
Perturbation techniques were used in the study of spherical and cylindrical gravitating sources \cite{6, 7}
and the effects of $f(R,T)$ gravity on astrophysical structures like wormholes and gravitational
vacuum stars were measured in \cite{8}.
Zubair et al. \cite{9} studied cosmic evolution in the presence of collision matter and measured its impacts on late-time dynamics
in the context of this viable modified theory.
Singh et al. \cite{10} proposed a new form for the parametrization of Hubble parameter which varies with the cosmic time $t$ and
studied the evolution of universe in the framework of $f(R,T)$ gravity.
The complexity of self-gravitating sources with different geometries has also been discussed in the realm of this theory \cite{11, 12}.
Keskin \cite{13} investigated oscillating scalar
field models, while Sofuoglu \cite{14} considered perfect fluid configuration and developed Bianchi type IX universe
model in $f(R,T)$ theory of gravity. The study of spacetimes with homogeneous and anisotropic features is important
to understand the the developments at early stages of the universe, which made the researchers to explore the
models for fluid configurations evolving under different scenarios in $f(R,T)$ theory of gravity \cite{15}-\cite{18}.

Anisotropy is an attractive feature of matter configurations which is
produced when the standard pressure is split into two different
contributions: the radial pressure and the transverse pressure
which are unlikely to coincide.
Due to its interesting behavior, it is still an active field of research.
Strong evidence shows that various interesting physical
phenomena give rise to a sufficiently large number of local
anisotropies for both type of regimes either it is of low or of extremely high density.
Among high density regimes, there are highly compact astrophysical objects
with core densities even higher than nuclear density ($3\times 10^{17} kg/m^3$). These objects have tendency to exhibit an anisotropic behavior
\cite{d1}, where the nuclear interactions are undoubtedly treated at relativistic levels.
Multiple factors which are considered responsible for the emergence of anisotropies in fluid pressure
include rotation, viscosity, the presence of a mixture of fluids of different types,
the existence of a solid core, the presence of a superfluid
or a magnetic field \cite{d2}. Even, some kind of phase transitions or pion condensation \cite{d3,d4}
can also be the sources of anisotropies.

In recent years, great interest has been developed for the construction of new analytic and anisotropic solutions for Einstein field equations,
which is not an easy task due to the highly nonlinear nature of the equations. In this context, Ovalle
\cite{d5} proposed an interesting and simple method named Gravitational Decoupling of Sources which applies a Minimal Geometric Deformation
(MGD) to the temporal and radial metric components together with a decoupling of sources, and constructs
new anisotropic solutions. Since its emergence, it has been used to develop a number of new anisotropic
solutions that are well-behaved and successfully represent stellar configurations.
Ovalle and Linares \cite{d6} worked out an exact solution for compact stars
and found its correspondence with the brane-world version of Tolman IV solution.
 In \cite{d7}, Ovalle
employed this approach to decouple the gravitational sources and constructed the
anisotropic solutions for spherical symmetry.
Furthermore, Ovalle and his collaborators \cite{d8} made an
effort to incorporate the effects of anisotropy in an isotropic solution and constructed three different versions of anisotropic
solutions from well known Tolman IV solution. Besides,
Contreras and Bargueno \cite{d9} considered $1 + 2$ circularly symmetric static spacetime
and constructed anisotropic solutions using MGD approach.
The Krori-Barua solution has also been extended to its anisotropic domain via this MGD approach, where
the effects of electromagnetic field has also been included \cite{d10}. Similarly, Durgapal-Fuloria perfect fluid solution has been considered
and its anisotropic version has been constructed \cite{d11}. Casadio et al. \cite{d12} employed this concept for the isotropization of an anisotropic
solution with zero complexity factor. Furthermore, the application of gravitational decoupling by MGD includes the solutions of
a variety of problems which are discussed in \cite{d13}-\cite{d32}, however the extended version
was developed in \cite{d33}. Sharif and Ama-Tul-Mughani \cite{d34} employed extended version of geometric
deformation decoupling method and developed the solution for anisotropic static sphere.
Furthermore, this method has been used to develop new black hole solutions
\cite{d35, d36}, and some other cosmological applications can be found in \cite{d37}-\cite{d41}.

Modified theories have also been examined in the context of MGD approach.
Sharif and Saba \cite{d42} obtained viable anisotropic solutions through this procedure in modified gravity.
Sharif and Waseem \cite{d43, d44} devoted their study to obtain spherically symmetric anisotropic and charged anisotropic
solution in modified theory of gravity.
Sharif and Ama-Tul-Mughani \cite{d45} computed exact charged isotropic as well as anisotropic solutions in a cloud
of strings. Maurya et al. \cite{d46} devoted their efforts to explore the possibility of getting solutions for
ultradense anisotropic stellar systems with the help of gravitational decoupling via MGD approach within the framework of $f(R,T)$ theory of gravity. However,
 decoupled gravitational sources by MGD approach in the framework of Rastall gravity have been presented in \cite{d47}.
Besides, the charged isotropic Durgapal-Fuloria solution in the framework of $f(R,T)$ theory has been investigated and extended it
 to achieve its anisotropic version \cite{d48}.

Inspiring by the work mentioned above, we have considered Tolman VII perfect fluid solution and extended it to its
an anisotropic version via gravitational decoupling by MGD in the $f(R,T)$ theory of gravity.
This paper has been arranged as follows: In the next section we present the
field equations of $f(R,T)$ theory of gravity in generic manner. Section III explains
MGD approach and splits the field equations into two less complex sectors. Section VI consists on junction conditions that establish a
relationship between inner and outer geometries. Section V represents the Tolman VII known solutions in $f(R,T)$ gravity and some useful results obtained by applying matching conditions. Section VI consists on new anisotropic solution in the context of $f(R,T)$ theory, which is followed by physical analysis of the proposed solution.
Last section concludes our main results.

\section{Gravitational decoupled field equations in $f(R,T)$ Gravity}

We consider the modified action for the $f(R,T)$ theory of gravity with $L_{m}$ as matter lagrangian and  $g$ as determinant of metric tensor $g_{ij}$.
Here, we consider $L_{\phi}$ as Lagrangian density for the new sector.
The total action for this theory will take the form as
\begin{eqnarray}\label{1}
I=\int f(R, T)\sqrt{-g}d^{4}x+\int L_{m}\sqrt{-g}d^{4}x+\beta\int L_{\phi}\sqrt{-g}d^{4}x,
\end{eqnarray}
which yields the following set of field equations by taking variation with respect to metric tensor
\begin{eqnarray}\label{2}
f_R R_{i j}-\frac{1}{2}fg_{i j}+(g_{i j}\Box-\nabla_{i}\nabla_{j})f_R=(1-f_T)
T^m_{i j}+\alpha\phi_{ij}-f_T\Theta_{i j}.
\end{eqnarray}
Here, $f_R=\partial f/\partial R,~f_{T}=\partial f/\partial T,~\square=g^{i j}\nabla_{i} \nabla_{j}$, whereas $\nabla_{i}$ represents the covariant derivative
and $\Theta_{ij}$ takes the form as
\begin{eqnarray}\label{th}
\Theta_{ij}=-2T_{ij}-P g_{ij}.
\end{eqnarray}
Further, the expression given in Eq.(\ref{2}) can be rewritten as
\begin{eqnarray}\label{3}
G_{ij}=R_{ij}-\frac{1}{2}R g_{ij}=T_{i j}^{(total)},
\end{eqnarray}
where
\begin{eqnarray}\label{e}
T_{ij}^{(total)}=\frac{1}{f_R}\left[(1+f_T)T_{ij}^{m}+ T_{ij}^{D}+\beta \phi_{ij}\right].
\end{eqnarray}
Here, the first term $T_{ij}^{m}$ represents energy momentum tensor whose mathematical expression assume the following form
\begin{eqnarray}\label{4}
T_{ij}^{m}=(\rho+P)U_{i}U_{j}+Pg_{ij},
\end{eqnarray}
where $U_{i}$ represents four velocity, while $\rho$ and $P$ indicate energy density and pressure, respectively.
The term $\phi_{ij}$ given in Eq.(\ref{e}) describes the extra source coupled with gravity through coupling constant $\beta$
which may offer some new fields and introduce anisotropy in stellar configurations.
However, the middle term in Eq.(\ref{e}) can be described mathematically as
\begin{eqnarray}\label{t}
T_{ij}^{D}=(\frac{f-R f_R}{2})g_{ij}+(\nabla_{i}\nabla_{j}-g_{ij}\Box)f_R+P g_{ij}f_T.
\end{eqnarray}

In this work, we have considered spherically symmetric static spacetime to
describe the interior configuration of a self-gravitating object which
can be defined as
\begin{eqnarray}\label{5}
ds^2=-e^{\xi(r)}dt^2+e^{\nu(r)}dr^2+r^2d\Omega^2,
\end{eqnarray}
where $d\Omega^2=d \theta^2+\sin^2 \theta d\phi^2$. With the consideration of above metric,
the set of field equations given in Eq.(\ref{3}) provides the following expressions
\begin{eqnarray}\label{e11}
(1+f_T)\rho+\beta\phi_{0}^{0}&=&e^{-\nu}\bigg[\bigg(\frac{\xi^{''}}{2}+\frac{\xi^{'}}{r}-\frac{\xi^{'}\nu^{'}}{4}+\frac{(\xi^{'})^2}{4}\bigg)f_{R}
-\bigg(\frac{2}{r}-\frac{\nu'}{r}\bigg)f_{R}'-f_{R}''\bigg]-\frac{f}{2}-P f_T,\\\label{e12}
P-\beta \phi_{1}^{1}&=&e^{-\nu}\bigg[\bigg(\frac{\nu'}{r}+\frac{\xi' \nu'}{4}-\frac{(\xi')^2}{4}-\frac{\xi''}{2}\bigg)f_{R}+\bigg(\frac{\xi'}{2}+\frac{2}{r}\bigg)f_{R}'\bigg]+\frac{f}{2},\\\label{e13}
P-\beta \phi_{2}^{2}&=&e^{-\nu}\bigg[\bigg(\frac{\nu'}{2r}-\frac{\xi'}{2r}-\frac{1}{r^2}+\frac{e^{\nu}}{r^2}\bigg)f_{R}-
\bigg(\frac{\nu'-\xi'}{2}-\frac{1}{r}\bigg)f_{R}'+f_{R}''\bigg]+\frac{f}{2},
\end{eqnarray}

Here, it is important to mention that prime denotes the derivative with respect to radial component $r$. The divergence of energy momentum tensor $T_{ij}^{(total)}$
yields the following expression
\begin{eqnarray}\label{7}
P'+(\rho+P)\frac{\nu'}{2}-\beta (\phi_{1}^{1})'+\frac{\xi'\beta}{2}(\phi_{0}^{0}-\phi_{1}^{1})+\frac{2\beta}{r}(\phi_2^2-\phi_1^1)=
\frac{1}{1+f_{T}}\bigg[(\rho+P)f_{T}'+(\rho'+\frac{3P'}{2})f_{T}\bigg].
\end{eqnarray}
By setting $\beta=0$, above expression can be reduced to the standard form for divergence of energy momentum tensor in $f(R,T)$ gravity
which is non-conserved.

In order to present our results meaningfully, we need to choose a particular $f(R, T)$ model.
The model which is in accordance with local gravity tests and satisfies cosmological constraints is
of fundamental importance as it offers a suitable substitute to dark source entities.
First, we consider a viable and cosmologically consistent model which is
linear combination of extended Starobinsky $f(R)$ model \cite{estar} and trace of energy momentum tensor $T$, i.e.
\begin{eqnarray}\label{10}
f(R, T)=R+\alpha R^2+ \gamma R^{n}+\lambda T,
\end{eqnarray}
where $\alpha, \gamma \in \Re$ and $n\geq 3$. As we can see that the model under consideration
is of the form $f(R,T)= f(R)+\lambda T$. Here, $\lambda T$ presents corrections to $f(R)$ model, while the choice of $\alpha=\gamma=0$ offers the $\lambda$ corrections to
GR. The construction of extended Starobinsky model is based on the inclusion of $n$th-order curvature term in Starobinsky model \cite{star}, where $n\geq 3$. This inclusion makes it able to incorporate effects of higher order curvature contributions.
It satisfies both viability criteria as it secures positivity of $1st$ and $2nd$-order derivatives with respect to Ricci scalar $R$.
Due to its viability and consistency, it has been used extensively in literature. After the consideration of this $f(R,T)$ model, the system of equations (\ref{e11})-(\ref{e13})
assumes the following form
\begin{eqnarray}\label{e21}
\tilde{\rho}+F_1&=&\bigg[\frac{1}{r^2}+e^{-\nu}\bigg(\frac{\nu'}{r}-\frac{1}{r^2}\bigg)\bigg]f_{R},\\\label{e22}
\tilde{P}_r+F_2&=&\bigg[e^{-\nu}\bigg(\frac{\xi'}{r}+\frac{1}{r^2}\bigg)-\frac{1}{r^2}\bigg]f_{R},\\\label{e23}
\tilde{P}_t+F_3&=&e^{-\nu}\bigg[\frac{\xi''}{2}-\frac{\nu'}{2r}+\frac{\xi'}{2r}-\frac{\xi'\nu'}{4}+\frac{(\xi')^2}{4}\bigg]f_{R},
\end{eqnarray}
where
\begin{eqnarray}\label{14*}
\tilde{\rho}&=&\rho+\frac{\lambda}{2}(3\rho-P)+\beta \phi_0^0,\\\label{14**}
\tilde{P}_r&=& P-\frac{\lambda}{2}(\rho-3P)-\beta\phi_1^1,\\\label{14***}
\tilde{P}_t&=& P-\frac{\lambda}{2}(\rho-3P)-\beta\phi_2^2.
\end{eqnarray}

In equations (\ref{e21})-(\ref{e23}), $F_1$, $F_2$ and $F_3$ are the terms containing curvature terms of different orders and its derivatives.
We can see that the system of equations (\ref{e11})-(\ref{e13}) is indefinite and contains seven unknowns, i.e., $\rho,~P $ (physical parameters), $\xi,~\nu$ (metric potentials)
and $\phi_0^0$, $\phi_1^1$, $\phi_2^2$ (independent components of extra source).
% Before moving towards next section, first we define physical parameters,
%total energy density ($\bar{\rho}$) and total principal stresses ($\bar{P_r}$, $\bar{P_t}$)
%in the following manner
%\begin{eqnarray}\label{14*}
%\bar{\rho} &=& \rho^{eff}+\beta \phi_0^0,\\\label{14**}
%\bar{P_r} &=& P^{eff}-\beta\phi_1^1,\\\label{14***}
%\bar{P_t}&=&P^{eff}-\beta \phi_2^2,
%\end{eqnarray}
%where effective quantities are given in Eqs.(\ref{e11*})-(\ref{e12*}).
In Eqs.(\ref{14*})-(\ref{14***}), we can clearly see that the additional source $\phi_{ij}$ is responsible for the introduction of
anisotropy in a self-gravitating system and for the construction of anisotropic parameter
 which takes the following form for the case under consideration
\begin{eqnarray}\label{15}
\tilde{\Delta}=\tilde{P}_t-\tilde{P}_r=\beta(\phi_1^1-\phi_2^2).
\end{eqnarray}
The system of Eqs.(\ref{e11})-\ref{e13}) may undoubtedly be treated as an anisotropic fluid, which would make it possible to
evaluate five unknown functions which include three total thermodynamical observables given in set of equations (\ref{14*})-(\ref{14***})
and two metric potentials $\xi(r)$ and $\nu(r)$. To achieve our target, we need to evaluate these unknowns.
Thus, a systematic approach proposed by Ovalle \cite{d5}
is adopted.
%This scenario leads to the complex situation regarding the combination of metric potentials.
\section{MGD Approach}

It is a hard task to find the analytic solutions for a system of equations which is highly non-linear. In order to tackle the situation,
 we adopt a simple but strong technique termed as gravitational decoupling via MGD approach, which opens
 a new window in the construction of exact solutions of non-linear system of field equations. Following the methodology, we divide the system with complex
 gravitational sources into two individual sectors and evaluate their corresponding solutions. The combined effects of the solutions
 lead towards the successful construction of anisotropic solution. The division of complex system takes place in such a way that
 field equation related to the source term $\phi_{ij}$ appears in a separate sector.

 First, we consider spherically symmetric spacetime with line element given below
\begin{eqnarray}\label{16}
ds^2=-e^{\mu(r)}dt^2+\frac{dr^2}{\eta(r)}+r^2d\Omega^2.
\end{eqnarray}
where $\eta(r)=1-\frac{2m}{r}$ and $m$ represents mass function with radial dependence. Next, we consider
 MGD transformation for metric coefficients ($\mu,~\eta$) which is defined as
\begin{eqnarray}\label{17}
\mu \mapsto \xi&=&\mu+\beta k, ~~ \eta \mapsto e^{-\nu}=\eta+\beta h,
\end{eqnarray}
where $h$ and $k$ describe radial and temporal deformations respectively and both of these functions have only radial dependence.
The MGD technique leads to the following conditions
\begin{eqnarray}\label{18}
k\mapsto 0, ~~~ h \mapsto h^{\star},
\end{eqnarray}
Here, we can easily see that the deformation has been introduced in radial component only,
 whereas temporal coefficient is not exposed to any change. Resultantly, we are able to extract following expressions
\begin{eqnarray}\label{19}
\xi\mapsto \mu &=&\xi(r),~~~\eta\mapsto e^{-\nu}=\eta(r)+\beta h^{\star}(r).
\end{eqnarray}
Insertion of Eq.(\ref{19}) into Eqs.(\ref{e21})-(\ref{e23}) will lead
to a very complex situation where it is very difficult to separate the both sectors, one corresponds to the
$f(R,T)$ modified equations, while the other is related to the additional source $\phi_{ij}$. In this situation, we have two possibilities:
Either we will apply MGD on right hand side of the system (\ref{e21})-(\ref{e23}) only or we will choose $\alpha=\gamma=0$ in Eq.(\ref{10}) and then apply MGD.
The first possibility is not a good choice mathematically. Thus we consider the second possibility which leads to the situation where
$F_1=F_2=F_3=0$ and $f_R=1$ in the Eqs.(\ref{e21})-(\ref{e23}). After this consideration, we apply MGD on the system Eqs.(\ref{e21})-(\ref{e23})
which is split into two sectors. Here, the first set corresponds to the isotropic fluid configuration and is obtained by assuming $\beta=0$.
Resultantly, we get the following mathematical expressions which represent modified $f(R,T)$ gravity system
\begin{eqnarray}\label{20}
\rho+\frac{\lambda}{2}(3\rho-P)&=&\frac{1}{r^2}-\frac{\eta}{r^2}-\frac{\eta'}{r^2},\\\label{20*}
P-\frac{\lambda}{2}(\rho-3P)&=&\eta\bigg(\frac{\xi'}{r}+\frac{1}{r^2}\bigg)-\frac{1}{r^2},\\\label{20**}
P-\frac{\lambda}{2}(\rho-3P)&=&\eta\bigg(\frac{\xi''}{2}+\frac{(\xi')^2}{4}+\frac{\xi'}{2r}\bigg)+\frac{\xi'\eta'}{4}+\frac{\eta'}{2r},
\end{eqnarray}
which further can be re-written as
\begin{eqnarray}\label{e1}
 \rho &=& \frac{1}{(1+\lambda)(1+2\lambda)r^2}\left[(1+\frac{3\lambda}{2})(1-\eta'r-\eta)+\frac{\lambda}{2}(\eta\xi'r+\eta-1)\right],\\\label{e2}
 P &=& \frac{1}{(1+\lambda)(1+2\lambda)r^2}\left[(1+\frac{3\lambda}{2})(\eta\xi'r+\eta-1)+\frac{\lambda}{2}(1-\eta'r-\eta)\right].
\end{eqnarray}
Here, isotropy equation $G_{11}=G_{22}$ leads to the following condition
\begin{eqnarray}\label{iso1}
 4(1-\eta)+2r(\eta'-\nu\xi')+r^2(2\eta\xi''+\eta\xi'+\eta'\xi') &=&  0,
\end{eqnarray}
Here, we can see that all the solutions obtained in GR are also present in the modified $f(R,T)$ gravity. However, it is important to note that
there is a quantitative difference between the Einstein's theory
of gravity and the modified $f(R,T)$ theory of gravity
when one considers the geometric and material content.
 In fact, the two models share only the geometrical content but
not the material one. In this situation, any solution to
Einstein theory of GR can be regarded as a solution in modified $f(R,T)$ system.
The conservation equation for the above system takes the form as
\begin{eqnarray}\label{21}
P'+(\rho+P)\frac{\eta'}{2}&=&\frac{1}{1+\lambda}(\rho'+\frac{3P'}{2}).
\end{eqnarray}
Here, the terms on the right side of the equations are
additional which appear due to the $f(R,T)$ gravity and are coupled with
the help of coupling constant $\lambda$. It is of the fundamental importance because
thermodynamical behavior of physical quantities depend on the value of $\lambda$.
Under the assumption $\lambda = 0$, all the above expressions are reduced to
the standard GR field equations for isotropic matter distribution \cite{d25}.
 The second set of equations corresponds to the additional gravitational source $\phi_{ij}$  and assumes the following form
\begin{eqnarray}\label{22}
\phi^{0}_{0}&=&-\left(\frac{{h^{\star}}'}{r}+\frac{h^{\star}}{r^2}\right),\\\label{22*}
\phi^{1}_{1}&=&-\frac{h^{\star}}{r}\Big(\frac{1}{r}+{\xi}'\Big),\\\label{22**}
\phi^{2}_{2}&=&-\left[\frac{h^{\star}}{4}\Big(2{\xi}''+{{\xi}'}^{2}+\frac{2{\xi}'}{r}\Big)+\frac{(h^{\star})'}{4}
    \Big({\xi}'+\frac{2}{r}\Big)\right],
\end{eqnarray}
with conservation equation
\begin{eqnarray}\label{23}
(\phi_1^1)'+\frac{\xi'}{2}(\phi_0^0-\phi_1^1)-\frac{2}{r}(\phi_2^2-\phi_1^1)&=&0.
\end{eqnarray}
The Eqs.(\ref{21}) and (\ref{23}) clearly show that both the systems given in Eqs.(\ref{20})-(\ref{20**}) and (\ref{22})-(\ref{22**})
conserve independently and interact gravitationally. The combined solution of Eqs.(\ref{21}) and (\ref{23}) leads to the the general conservation
equation for total energy momentum tensor in $f(R, T)$ gravity.
Now it is worthwhile to mention here that from now onward
we will characterize the total physical quantities ($\bar{\rho}$, $\bar{P_r}$ and  $\bar{P_t}$)
in the following manner
\begin{eqnarray}\label{l}
\bar{\rho}&=&\rho+\beta\phi^{0}_{0},\\\label{l*}
\bar{P_{r}}&=&P-\beta\phi^{1}_{1},\\\label{l**}
\bar{P_{t}}&=&P-\beta\phi^{2}_{2},
\end{eqnarray}
where $\rho$ and $P$ are given in Eqs.(\ref{e1})-(\ref{e2}). These equations contain extra geometric
terms which is coupled via coupling parameter $\lambda$. It is also important to note that anisotropic factor given in Eq.(\ref{15}) and anisotropic
factor obtained from equations (\ref{l*}) and (\ref{l**}) are completely same.
Moreover, the inner geometry of the MGD model is defined by the following line element
\begin{eqnarray}\label{24}
ds^{2}=-e^{\mu(r)}dt^{2}+\frac{dr^{2}}{h^{\star}(r)}+r^{2}(d\theta^{2}+\sin^{2}\theta).
\end{eqnarray}

\subsection{Junction Conditions}

Junction conditions are of fundamental importance in study of stellar objects. The smooth matching of interior and exterior geometries
at the surface of stellar configurations is required to ensure well-behaved compact structures. Here, it is important
to mention that that Israel-Darmois (ID) matching conditions do work in the case of $f(R, T)=R+\lambda T$ in the same way as these work in GR, i.e., the appropriated
outer spacetime corresponds to Schwarzschild geometry. This is so because the modification introduced in matter source represented by $T$ is vanished across
the boundary. Thus, one can join the interior geometry with Schwarzschild outer spacetime by imposing the first and second fundamental forms at the boundary of the star.
Here, the inner geometry of gravitating source via MGD approach is given by
\begin{equation}\label{26}
    ds^{2}=-e^{\xi(r)_{-}}dt^2+\Big(1-\frac{2\tilde{m}(r)}{r}\Big)^{-1}dr^{2}+r^{2}(d\theta^{2}+\sin^{2}\theta d\phi^{2}),
\end{equation}
where $\tilde{m}=m(r)-\frac{\beta}{2}h^{\star}(r)$ and describes interior mass. For outer spacetime, we consider general
exterior spherically symmetric line element which is defined as
\begin{equation}\label{27}
    ds^{2}=- e^{\xi(r)^{+}}dt^2+e^{\nu(r)_{+}}dr^{2}+ r^{2}(d\theta^{2}+\sin^{2}\theta d\phi^{2}).
\end{equation}
 Following the continuity of first fundamental form at the boundary where interior and exterior geometries coincide (i.e., $r=\mathcal{R}$),
 we obtain the following results
\begin{equation}\label{28}
    \xi(r)^{+}= \xi(r)^{-}~~  and ~~ 1-\frac{2M_{o}}{\mathcal{R}}+ \beta h^{\star}(\mathcal{R})= e^{-\nu(\mathcal{R})^{-}}.
\end{equation}
These actually follow from the conditions $[ds^{2}]_{\sum}=0$ with $V_{\sum}=V(\mathcal{R})^{+}-V(\mathcal{R})^{-}$, where $V=V(\mathcal{R})$ be an arbitrary function.
Here, $h^{\star}(\mathcal{R})$ has been used to represent the total deformation via MGD scheme, while $M_{0}=m(\mathcal{R})$ appears for mass at the boundary of stellar configuration.
The continuity of second fundamental form finds its origin in the hypothesis $[T^{D}_{\xi\eta}X^{n}]=0$, where $X^{n}$ indicates a unit four vector in radial direction.
Following the aforementioned fundamental form, we find
\begin{equation}\label{29}
\bar{P}_{\mathcal{R}}=P_{\mathcal{R}}-\beta(\phi^{1}_{1}(\mathcal{R}))^{-}=-\beta(\phi^{1}_{1}(\mathcal{R}))^{+},
\end{equation}
which further leads to
\begin{equation}\label{30}
P_{\mathcal{R}}+ \frac{\beta h^{\star}(\mathcal{R})}{\mathcal{R}}\Big({\xi}'+\frac{1}{\mathcal{R}}\Big)=
\frac{\beta k^{\star}(\mathcal{R})}{\mathcal{R}}\Big(\frac{1}{\mathcal{R}}+\frac{2M}{\mathcal{R}(\mathcal{R}-2M)}\Big),
\end{equation}
where the choice of $k^{\star}$ provides ``radial geometric deformation'' of Schwarzschild metric, i.e.,
\begin{equation}\label{31}
    ds^{2}=-\Big(1-\frac{2M}{r}\Big)dt^2+ \Big(1-\frac{2M}{r}+\beta k^{\star}\Big)^{-1} dr^{2}+r^{2}(d\theta^{2}+\sin^{2}\theta d\phi^{2}).
\end{equation}
Thus, we have necessary and sufficient conditions for the smooth matching of interior MGD metric and a deformed Schwarzschild metric. Next we consider standard "Schwarzschild metric" in order to represent outer spacetime, which is here based on the assumption $k^{\star}=0$.
 Resultantly, we have
\begin{equation}\label{32}
 \bar{P}_{\mathcal{R}}=P_{\mathcal{R}}-\beta\phi_1^1(\mathcal{R})=P_{\mathcal{R}}+ \frac{\beta h^{\star}(\mathcal{\mathcal{R}})}{\mathcal{R}}\Big(\xi'+\frac{1}{\mathcal{R}}\Big) =0.
\end{equation}
The last equation has important consequences as the
compact object will be in equilibrium in an exterior
spacetime without material content only if the total radial pressure at the boundary vanishes.
This makes us that the last condition (\ref{32}) determines the size of the compact object, i.e., the radius $\mathcal{R}$.
Thus, it can be concluded that the material content is confined within
the region $0\leq r \leq \mathcal{R}$.

\subsection{Anisotropic Interior Solutions}

Here, we consider the well-known Tolman VII solution which describes a perfect fluid stellar configuration \cite{tol}.
\begin{eqnarray}\label{33}
    e^{\xi(r)}&=&B^{2}\sin ^2 z, \\\label{33*}
    e^{-\nu(r)}&=&\eta(r)=1-\frac{r^2}{\mathcal{R}^2}+\frac{4r^4}{A^4},
 \end{eqnarray}
where $z=\log{\sqrt{\frac{e^{-\frac{\nu(r)}{2}}+\frac{2r^2}{A^2}-\frac{A^2}{4\mathcal{R}^2}}{C}}}$. Moreover, inside the stellar configuration,
we have thermodynamical quantities in the following form
\begin{eqnarray}\label{36*}
\rho&=&\frac{1}{r^2(1 +\lambda)(1+2\lambda)}\bigg[1+\lambda-r(2+3\lambda)b_1b_1'-
 b_1^2(1+\lambda-r\lambda z'\cot{z})\bigg],\\\label{36**}
P&=& \frac{-1}{r^2(1 +\lambda)(1+2\lambda)}\bigg[1+\lambda+r\lambda b_1b_1'-
 b_1^2(1+\lambda+r(2+3\lambda)z'\cot{z})\bigg],\\\nonumber
\end{eqnarray}

where $b_1=\sqrt{1-\frac{r^2}{\mathcal{R}^2}+\frac{4r^4}{A^4}}$, while $z'$ and $b_1'$ represent radial derivatives of the expressions $z$ and $b_1$ respectively .

Here, the constants $A,~ B, ~C$ can be obtained through matching conditions between the interior and exterior solutions (Schwarzschild spacetime). This yields
 \begin{eqnarray}\label{35}
 A^{2}&=&\left(\frac{4\mathcal{R}^5}{\mathcal{R}-2M_0}\right)^{\frac{1}{2}},
\\\label{35*}
    B&=&\pm\csc\left[{\cot^{-1}\left(\frac{M_0b_2b_3}{\mathcal{R}^7}\right)}\right]b_2,
\\\label{35**}
  C&=&e^{-2\cot^{-1}\left(\frac{M_0b_2b_3}{\mathcal{R}^7}\right)}\left(2b_2-\frac{1}{2\mathcal{R}^2}\left(\frac{\mathcal{R}^5}{\mathcal{R}-2M_0}\right)^{\frac{1}{2}}\right),
 \end{eqnarray}
 where $b_2=\sqrt{1-\frac{2M_0}{\mathcal{R}}}$ and $b_3=\left(\frac{\mathcal{R}^5}{\mathcal{R}-2M_0}\right)^{\frac{3}{2}}$.
The Above solution ensures that continuity of exterior and interior region at surface of star may be changed in the presence of additional source ($\phi_{ij}$). To obtain anisotropic solution i.e., $\beta\neq0$, temporal as well as radial components are defined by Eqs.(\ref{19}), (\ref{33}) and (\ref{33*}).
The source term $(\phi_{ij})$ and $h^{\star}$ are connected by the relations given in Eqs.(\ref{22})-(\ref{22**}) which can be solved by applying additional constraints. In this scenario, we impose another condition in order to obtain an interior solution in the next subsection.
\subsection{Mimic constraint and anisotropic solution }

The compatibility of Schwarzchild exterior geometry and interior geometry of a gravitating source is dependent on the condition $\beta{\phi^{1}_{1}}(\mathcal{R}) \sim P(\mathcal{R})$ as it is very obvious from Eq.(\ref{32}). Thus, in order to obtain a physically and mathematically acceptable solution we can make the following simplest choice
\begin{equation}\label{36}
   \phi^{1}_{1}(r)=P(r),
\end{equation}
which establishes the equality between Eqs.(\ref{e2}) and (\ref{22*}). It provides us a general expression for the deformation
function $h^{\star}(r)$, i.e.,
\begin{eqnarray}\label{36*}
  h^{\star}(r) &=& \frac{(2+3\lambda)(1-\eta\mu'r-\eta)-\lambda(1-\eta'r-\eta)}{2(1+\lambda)(1+2\lambda)(1+r\mu')},
\end{eqnarray}
which defines the radial component with the help of Eq.(\ref{19}) as follows
\begin{equation}\label{37}
    e^{-\nu(r)}= \eta(r)+\beta\frac{(2+3\lambda)(1-\eta\mu'r-\eta)-\lambda(1-\eta'r-\eta)}{2(1+\lambda)(1+2\lambda)(1+r\mu')}.
\end{equation}
We can see from Eqs.(\ref{22})-(\ref{22**}) and (\ref{37}) that the Tolman VII solution is minimally deformed by $\phi_{ij}$ and is
ready to describes an anisotropic solution.
For $\beta=0$, Eq.(\ref{37}) is reduced to the standard spherical solution.
Here, the first fundamental form leads to the following result
\begin{eqnarray}\nonumber
    B^{2}\sin ^2 z\mid_{r=\mathcal{R}}=1-\frac{2M}{\mathcal{R}},
\\\label{38}
    \eta(r)+\beta\frac{(2+3\lambda)(1-\eta\mu'r-\eta)-\lambda(1-\eta'r-\eta)}{2(1+\lambda)(1+2\lambda)(1+r\mu')} \mid_{r=\mathcal{R}}=1-\frac{2M}{\mathcal{R}},
\end{eqnarray}
Moreover, the continuity of second fundamental form $\bar{P}(r)|_{r={\mathcal{R}}}=0$ (given in Eq.(\ref{32})) leads to
\begin{equation}\label{39}
(1-\beta)P(r)|_{r={\mathcal{R}}}=0.
\end{equation}
Eqs.(\ref{38}) and (\ref{39}) present the conditions that are inevitably required to satisfy at the boundary of a star.
In this situation, the physical quantities ($\bar{\rho}$, $\bar{P}_{r}$, $\bar{P}_{t}$) can assume the following forms
\begin{eqnarray}\nonumber
\bar{\rho}&=&\rho-\frac{\beta}{r^2\lambda_{11}\lambda_{12}Y_2^2}\bigg[\lambda_{11}+r^2(b_1'^2\lambda Y_2+
2\lambda_{11}(\csc^2{z}z'^2-z''\cot{z}))-b_1^2\lambda_1+rb_1^2(z'\lambda_{46}\cot{z}+Y_1)+rb_1^3\\\label{f1}&&(r\lambda b_1''(1+2rz'\cot{z})-
2b_1'(1+r(4z'\lambda_1\cot{z}+Y_1)))\bigg] \\\label{f2}
\bar{P}_r&=&(1-\beta)P,\\\label{f3}
\bar{P}_t&=&\bar{P}_r+\bar{\Delta}(r),
\end{eqnarray}
Here, we have renamed some expression for the sake of convenience as $\lambda_{ij}=(i+j\lambda)$ with $(i,j=1,2,3...)$,
$Y_1=r(-\lambda_{46}+\lambda_{45}\csc^2{z})z'^2+r\lambda z'' \cot{z})$ and $Y_2=1+2rz' \cot{z}$.
However, the anisotropic factor $\bar{\Delta}$ takes the form as
\begin{eqnarray}\nonumber
\bar{\Delta}(r)&=&\frac{1}{r^2\lambda_{11}\lambda_{12}}\bigg[\frac{r}{2Y_2^2}\bigg\{2Y_2Y_3(z'\cot{z}-rz')+
r\cot{z}z''-(1+rz'\cot{z})(2Y_3Y_4-Y_2(r\lambda b_1'^2+b_1(-b_1'(\lambda_{21}+
2 r\lambda_{23}\\\label{41}&&z'\cot{z})+r\lambda b_1''-\lambda_{23}b_1^2Y_4)))\bigg\}-\lambda_{11}-r\lambda b_1b_1'+
 b_1^2(\lambda_{11}+r\lambda_{23}z'\cot{z})\bigg],
\end{eqnarray}
where $Y_3=(\lambda_1+r\lambda b_1b_1'-b_1^2(\lambda_1+r\lambda_{23}z'\cot{z}))$, and $Y_4=z'(\cot{z}-rz'\csc^2{z})+
rz''\cot{z}$.
%\begin{figure}
% \epsfig{file=dbar.eps, width=.4\linewidth,
%height=1.5in}\epsfig{file=dbar.eps, width=.4\linewidth,
%height=1.5in} \caption{\label{Fig.1} shows the graphs of physical quantities for the new anisotropic solution and Tolman V perfect fluid solution
%. Herein, we set $R=4$ and $M_0=1$ with $\beta=0.5$. In left panel curves represent $\rho$ (Solid), $\tilde{\rho}$ (Dashed) whereas right panel $P$ (solid), $\tilde{P_r}$ (Dashed) and $\tilde{P_t}$ (Dotted).}
%\end{figure}
\begin{figure}
\centering
\includegraphics[width=0.40\textwidth]{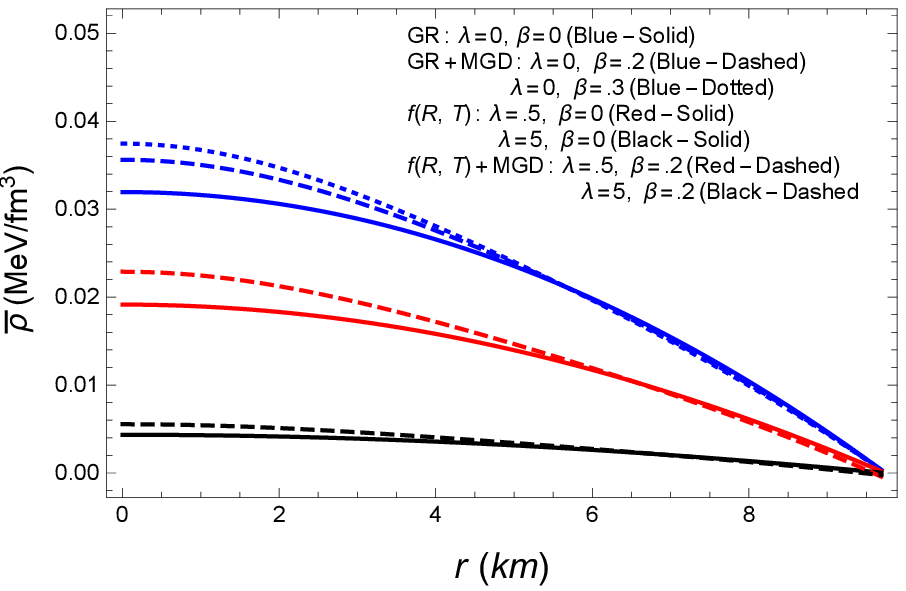}
\includegraphics[width=0.40\textwidth]{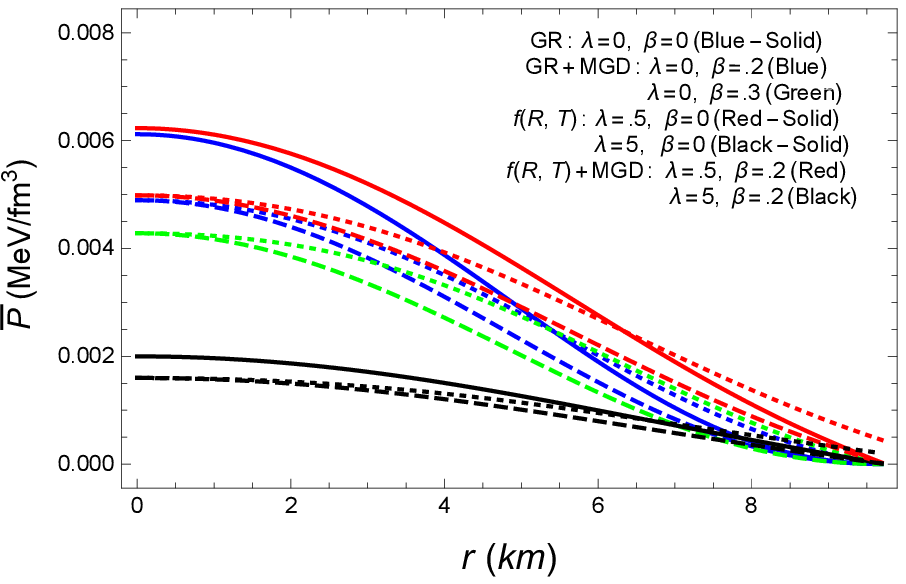}
\includegraphics[width=0.40\textwidth]{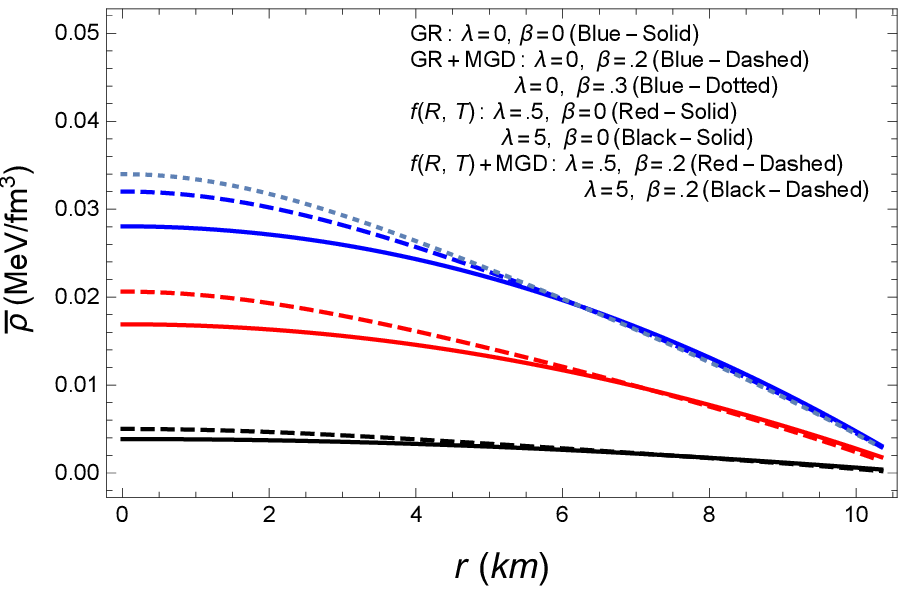}
\includegraphics[width=0.40\textwidth]{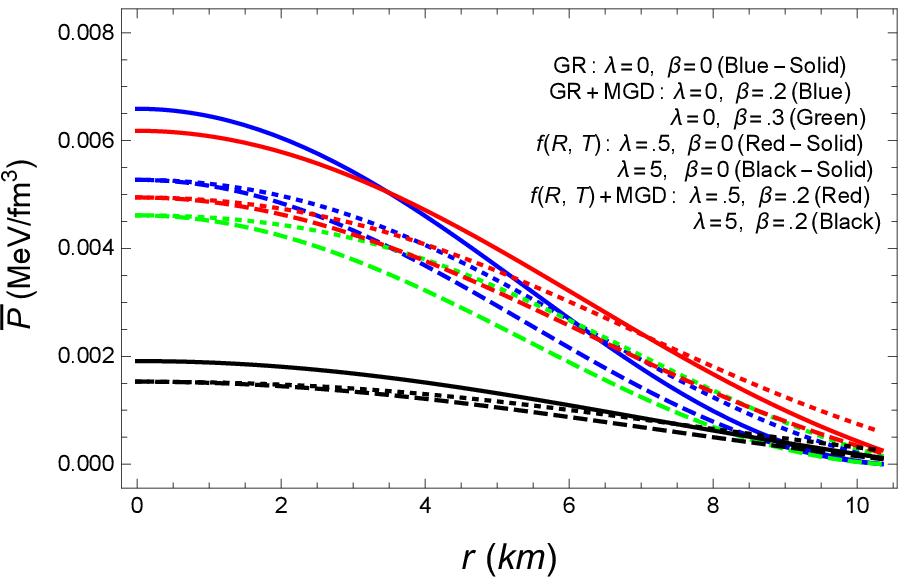}
\includegraphics[width=0.40\textwidth]{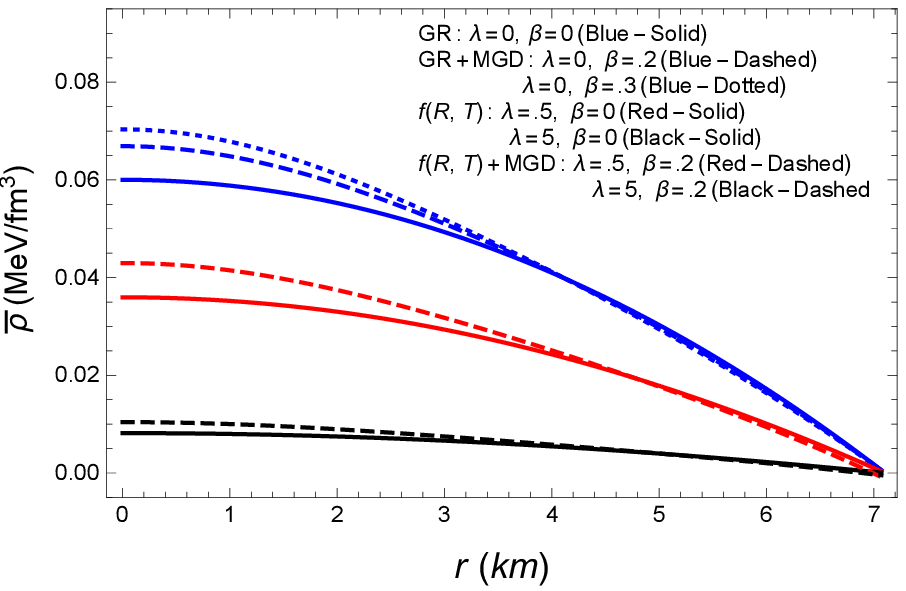}
\includegraphics[width=0.40\textwidth]{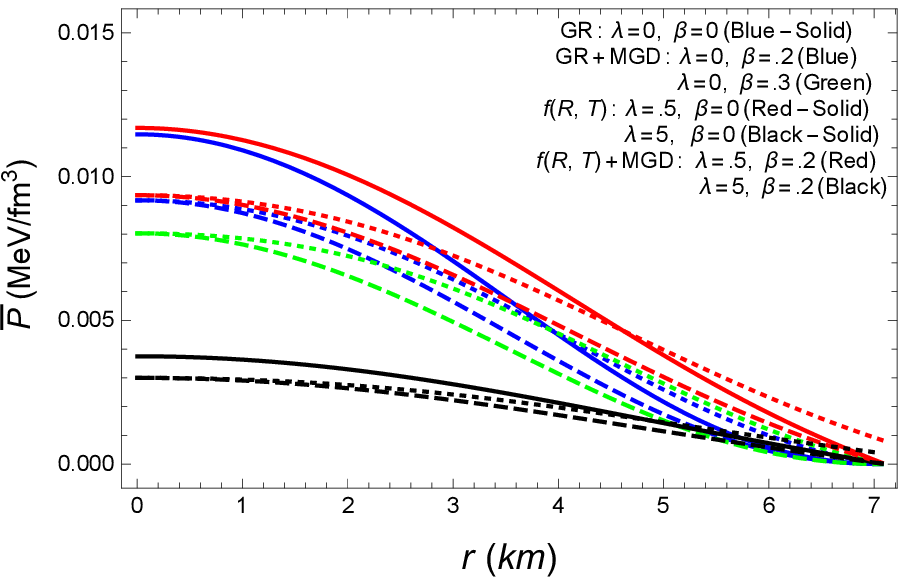}
\caption{ \scriptsize{Variation of energy density ($\bar{\rho}$) and pressure ($\bar{P}$) with respect to radial coordinate
 $r$ for compact stars, namely PSR J1614-2230 (1st Row), PSR 1937+21 (2nd row), SAX J1808.4-3658 (3rd row).
Left panel shows the evolution of energy density $\bar{\rho}$. In the right penal, the solid curves represent $P$, the dashed curves represent $\bar{P}_r$ and dotted curves represent
$\bar{P}_t$.}}
\label{fig1}
\end{figure}

\begin{figure}
\centering
\includegraphics[width=0.40\textwidth]{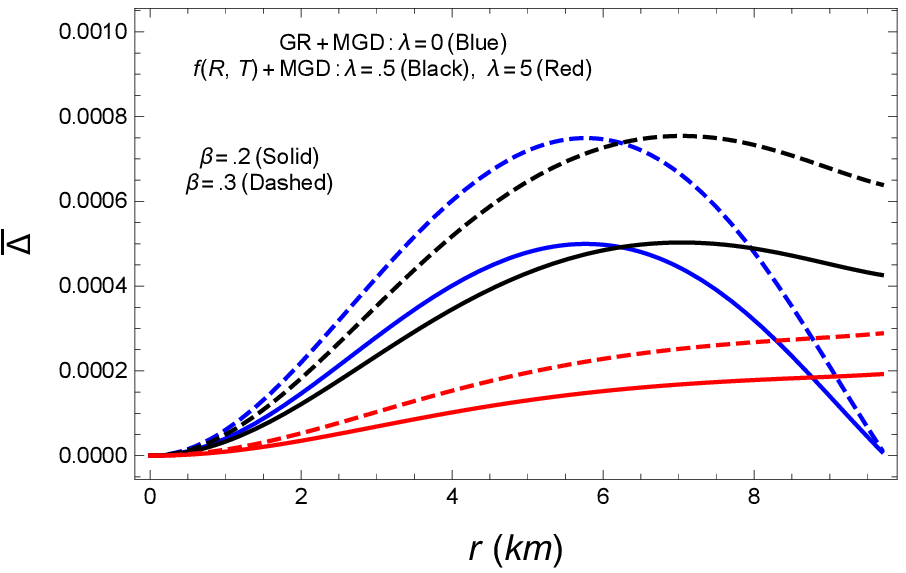}
\includegraphics[width=0.40\textwidth]{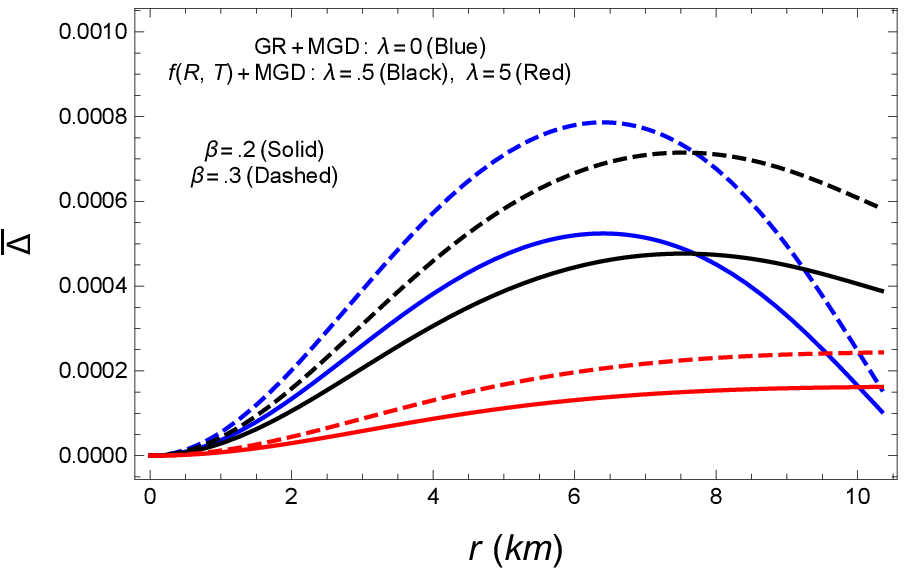}
\includegraphics[width=0.40\textwidth]{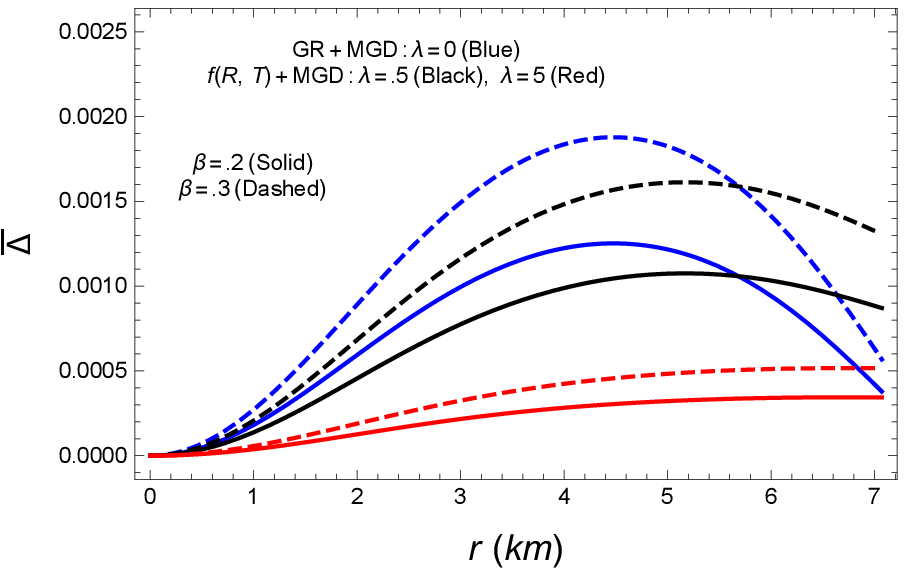}
\caption{\scriptsize{Variation of anisotropic factor for PSR J1614-2230 (1st Row, left penal), PSR 1937+21 (1st row, right penal), SAX J1808.4-3658 (2nd row)
with respect to radial coordinate $r$.}}
\label{fig2}
\end{figure}

\section{Physical Analysis of Anisotropic solution}
It is widely known that a theoretically well-behaved compact model
must satisfy some general criteria physically and mathematically in order to compare with the
astrophysical observation data. In the continuing section, we explore salient features of the above anisotropic solution which prove
quite helpful in the description of the structure of relativistic compact objects. For this purpose, we consider
three different compact star model, namely PSR J1614-2230, PSR 1937+21 and SAX J1808.4-3658, and present graphical analysis of these features
with the help of their observational data values. Thus, it will enable us to explore the suitability and capability of
new anisotropic solution for the characterization of the realistic structures. For graphical presentation, we have considered
four cases, i.e., GR, GR+MGD, $f(R,T)$ and $f(R,T)$+MGD, so that we can compare and discuss the results effectively.

\subsection{Thermodynamical variables}

The variation of thermodynamic observables which correspond to the total energy density ($\bar{\rho}$) and total pressure components ($\bar{P}_r$, $\bar{P}_t$),
has graphically been presented in Fig.\ref{fig1}. We can see that all these observables are in agreement with salient features of a well-behaved
stellar structure.
\begin{itemize}
\item The energy density ($\bar{\rho}$), total radial and transverse pressures ($\bar{P}_r$, $\bar{P}_t$) are positive definite in the whole domain.
  \item They attain their maximum values at center and then gradually decrease towards the boundary of the star.
  \item At the center, $\bar{P}_r=\bar{P}_t$ which corresponds to the zero anisotropy (i.e., $\bar{\Delta}=0$).
  \item At the boundary, radial pressure attains its minimum value which is close to the zero.
\end{itemize}
For graphical presentation,
we have fixed the values of $\lambda$ and observed that the values of thermodynamic parameters are decreased
as the value of coupling parameter $\lambda$ is increased.
\subsection{Anisotropic Factor}
The anisotropic factor ($\bar{\Delta}$) which is determined by the difference of radial and
transverse pressure facilitates greatly in the determination of the characteristics of a stellar structure.
Its value remain positive if $\bar{P}_r<\bar{P}_t$, which indicates towards a stable stellar configuration.
In Fig.\ref{fig2}, we can see the anisotropic factor has positive value throughout, while $\bar{\Delta}=0$ at the center and near the boundary surface it is decreased.
It can also been seen that as the value of $\beta$ is increased, anisotropy of the system is also increased. However, our anisotropic solution follows
all the salient features of a stable stellar structure when $0\leq\beta\leq 0.2$, otherwise the condition of attainment of maximum values of physical parameters at the center
is violated for $\beta> 0.2$.

\begin{figure}
\centering
\includegraphics[width=0.40\textwidth]{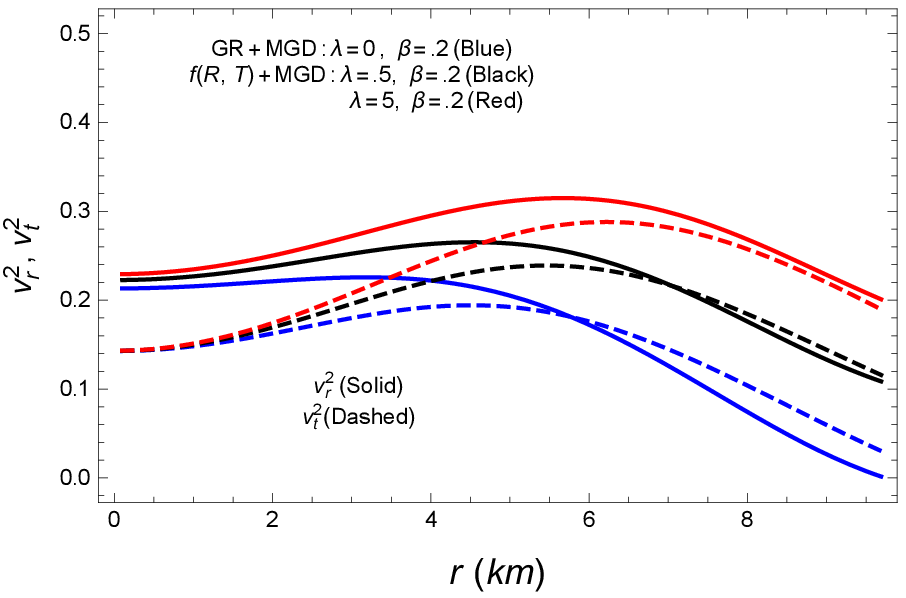}
\includegraphics[width=0.40\textwidth]{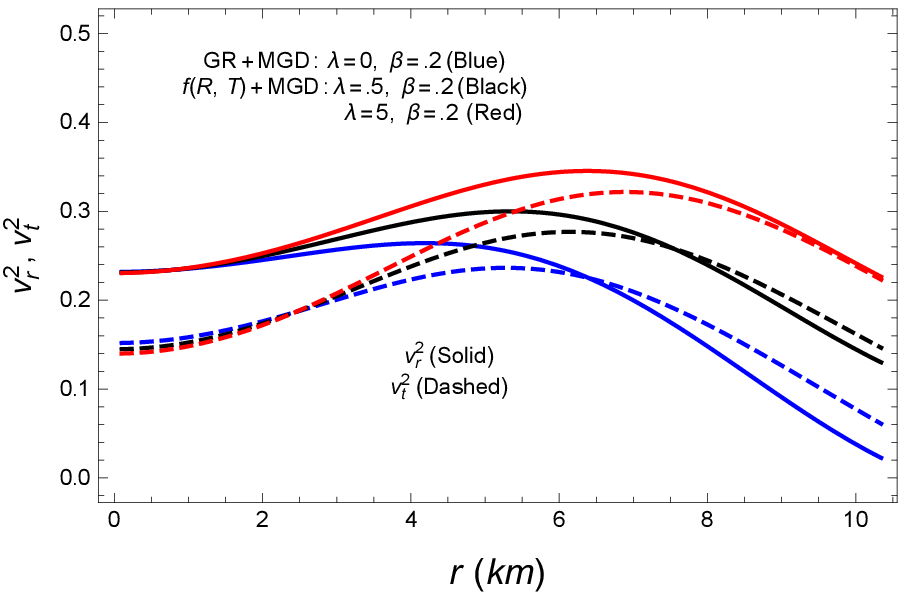}
\includegraphics[width=0.40\textwidth]{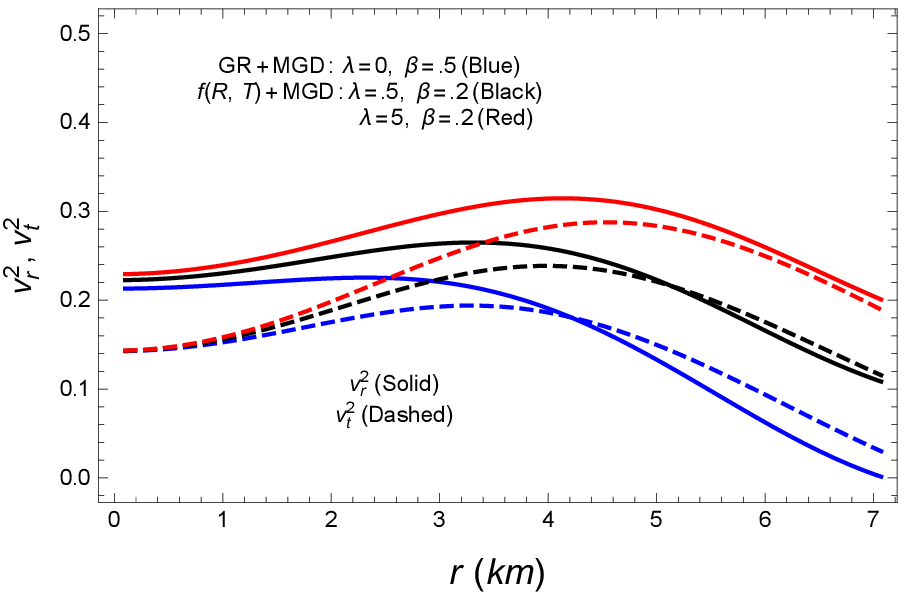}
\caption{ \scriptsize{Variation of $v_r^2$ (Solid) and $v_t^2$ (Dashed) for PSR J1614-2230 (1st Row, left penal),
PSR 1937+21 (1st row, right penal), SAX J1808.4-3658 (2nd row) with respect to radial coordinate $r$.}}.
\label{fig3}
\end{figure}
\begin{figure}
\centering
\includegraphics[width=0.40\textwidth]{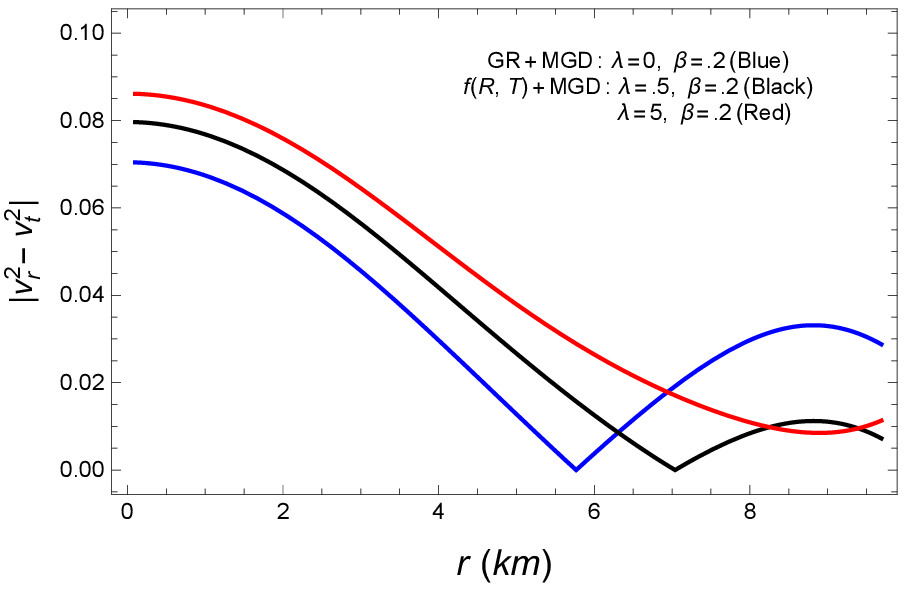}
\includegraphics[width=0.40\textwidth]{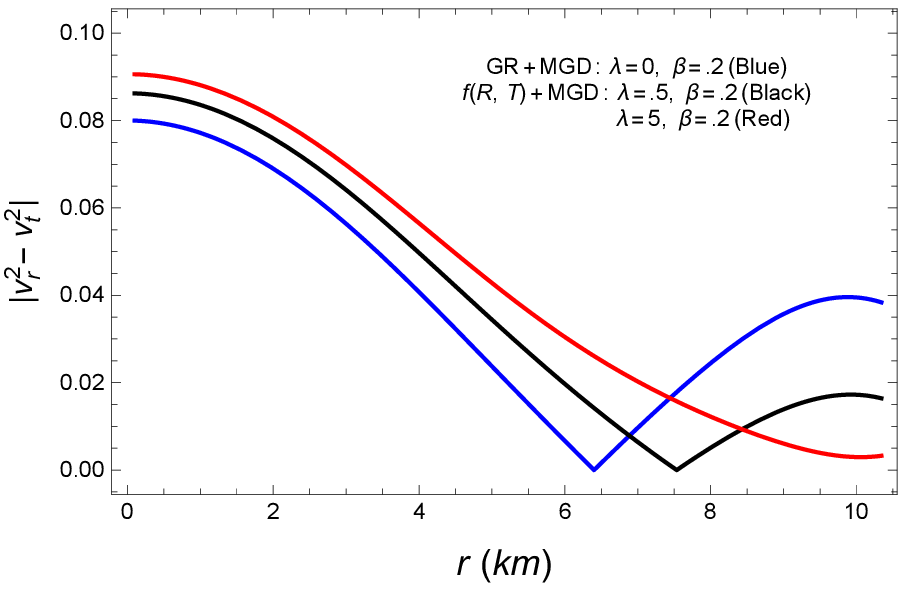}
\includegraphics[width=0.40\textwidth]{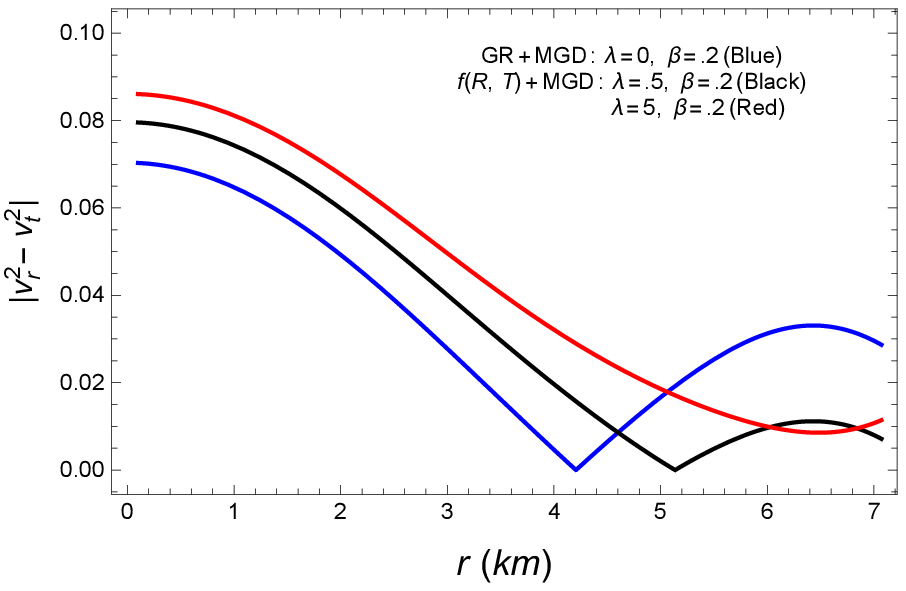}
\caption{ \scriptsize{Variation of $\mid v_t^2-v_r^2\mid $ for PSR J1614-2230 (1st Row, left penal),
PSR 1937+21 (1st row, right penal), SAX J1808.4-3658 (2nd row) with respect to radial coordinate $r$.}}.
\label{fig4}
\end{figure}
\subsection{Energy Conditions}
Energy conditions play a significant role in the identification of
normal and unusual nature of matter inside a stellar structure model. Thus, these
conditions have grasped much attention in the investigation of cosmological phenomena.
These are divided into four categories known as null energy condition (NEC), weak energy condition (WEC), strong energy condition (SEC) and
dominant energy condition (DEC). These conditions are satisfied if
the inequalities given below satisfy at every point inside the sphere

\begin{itemize}
  \item NEC: $\bar{\rho}+\bar{P_r}\geq 0$,  $\bar{\rho}+\bar{P_t}\geq 0$
  \item WEC: $\bar{\rho}\geq 0$, $\bar{\rho}+\bar{P_r}\geq 0$,  $\bar{\rho}+\bar{P_t}\geq 0$
  \item SEC: $\bar{\rho}+\bar{P_r}\geq 0$,  $\bar{\rho}+\bar{P_t}\geq 0$, $\bar{\rho}+\tilde{P_r}+2\bar{P_t}\geq 0$
  \item DEC: $\bar{\rho}-\mid\bar{P_r}\mid\geq 0$,  $\bar{\rho}-\mid\bar{P_t}\mid\geq 0$
\end{itemize}
In Fig.\ref{fig1}, it can clearly be seen that all the energy conditions are completely satisfied at every point
inside the sphere.

\subsection{Stability Analysis}
The stability of stellar structures occupies significant place in the scrutiny of physical
consistency of the models. In the recent years, different concepts have been used in order to gauge
the stability of compact star models.
The radial and transverse speed of sounds can mathematically be defined as
\begin{eqnarray}\label{v}
  v_r^2 &=& \frac{d \bar{P}_r}{d\bar{\rho}},\quad\quad v_t^2 =\frac{d \bar{P}_t}{d\bar{\rho}}.
\end{eqnarray}
For a physically stable structure, the causality condition must be satisfied which corresponds
to the situation that the values of radial ($v_r^2$) and transverse sound speed ($v_t^2$) must lie in the unit interval at each point inside
the stellar configuration. Moreover, Herrera and his co-researchers \cite{st1, st2} introduced another concept known as Herrera's cracking concept to
present the stability analysis of stellar models. This concept is based on the difference of radial and transverse speeds of sound
and claims that the regions where absolute value of this difference is less than $1$ are considered potentially stable,
i.e. $0\leq \mid v _r^2-v_t^2\mid\leq 1$. In the Figs.\ref{fig3} and \ref{fig4}, we can clearly assess that our stellar
model is potentially stable throughout the stellar distribution as it obeys both causality condition and Herrera's cracking concept.

\section{Concluding remarks}
The study the self-gravitating objects and construction of new stellar anisotropic models that
fulfills the basic and general requirements for a physically and mathematically admissible one is
an active field of research. In this context, the gravitational decoupling approach has proved an important development which provides
a simple but elegant method to construct exact interior solution of anisotropic relativistic objects.
This technique actually provides a mechanism to extend the isotropic spherical solutions to anisotropic ones
by inserting extra gravitational sources. Following this approach, we have developed anisotropic version of Tolman VII perfect fluid solution in $f(R,T)$
gravity. For this, we have taken into account the $f(R,T)$ model which is linear combination of extended starobinski model and trace of energy momentum tensor $T$,
i.e., $f(R,T)=R+\alpha R^2+\gamma R^n+\lambda T$ gravity, where $\alpha, \gamma \in \Re$, $n\geq 3$ and $\lambda$ is arbitrary positive constant.
In order to achieve our target, we introduced a new source in effective energy momentum tensor
and constructed corresponding field equations. Before the implication of MGD approach, we have chosen $\alpha=0=\gamma$, as we have found that
the presence of curvature terms of different orders and its derivatives leads to a complex situation where it becomes difficult to separate the
complex set of field equations into two less complicated sections. After this consideration,
we have followed the MGD approach and divided set of field equations into two sectors
which only interacts gravitationally. One corresponds to the
$f(R,T)$ field equations, while the other represents field equations for the extra source.
To establish the anisotropic solutions, we considered Tolman VII perfect fluid solution in $f(R,T)$ gravity
and evaluated constants appearing in the solution with the help of matching conditions.
We also used matching conditions to exhibit a relation between spherically symmetric interior metric
and exterior Schwarzschild spacetime. After having all the necessary ingredients, we implemented mimic
constraint for effective pressure, i.e., $\beta\phi_1^1(R)\sim P$,
on the behalf of matching condition given in Eq.(\ref{32}) and obtained anisotropic version of the solution under consideration.

In order to gauge the physical viability of the solution, we have presented detailed physical analysis of the solution.
taking into account three different compact star models, namely PSR J1614-2230, PSR 1937+21 and SAX J1808.4-3658. We have discussed the results
for four different cases, i.e., GR, GR+MGD, $f(R,T)$ and $f(R,T)$+MGD.
\begin{itemize}
  \item The graphical representation of thermodynamical variables has been provided in Fig.\ref{fig1}, where it can be seen that
all the variables preserve the salient features of a stable stellar configuration.
The behavior of total energy density with respect to to radial coordinate $r$
shows that as $r\rightarrow 0$, the total energy density attains its
maximum value for all the compact stars candidates. On the other side, the total radial pressure
and tangential pressure remain positive throughout inside the stellar configurations, whereas they attain same but maximum values
as $r\rightarrow 0$. Besides, both the features total energy density and total pressure components are monotonically decreasing functions
of $r$. Moreover, we have fixed the values of $\beta$,
whereas we have chosen two different values for $\lambda$ and observed that stellar model exhibit consistent
behavior near the boundary for both values of $\lambda$.
\item In order to visualize the effects of anisotropic factor $\bar{\Delta}$, we have chosen different values of $\beta$ and presented the results
graphically in Fig.\ref{fig2}, where we have seen that intensity factor $\beta$ and anisotropic factor
$\bar{\Delta}$ have direct relation. It has also been observed that our solution preserves all the salient features
of stable stellar configuration for $0\leq\beta\leq 0.2$, whereas tangential pressure does not show monotonically decreasing behavior for $\beta>0.2$.
\item It is very obvious from the Fig.\ref{fig1} that all the energy conditions, i.e.,
$\bar{\rho}\geq 0$, $\bar{\rho}+\bar{P}_r\geq 0$, $\bar{\rho}+\bar{P}_t\geq 0$,
$\bar{\rho}+\bar{P}_r+2\bar{P}_t\geq 0$
 $\bar{\rho}-\mid\bar{P_r}\mid\geq 0$, $\bar{\rho}-\mid\bar{P_t}\mid\geq 0$ are satisfied throughout
inside the stellar configuration which ensures physical viability of the stellar model.
\item The stability analysis shows that our stellar model obeys both causality condition
and Herrera's cracking concept which is very obvious from Figs.\ref{fig3} and \ref{fig4}.
Fig.\ref{fig3} shows that the radial and transverse speeds of sound lie
in the unit interval, while Fig.\ref{fig4} clearly shows that the difference of the squares of sound speeds
also lie in the unit interval. Both these conditions ensure
the consistency and stability of our solution.
\end{itemize}

Finally, we can safely claim that our proposed $f(R,T)$ model represents stellar structure systems which are physically
stable and is suitable to describe anisotropic nature of compact stars.
The results against $\lambda=0$  corresponds to GR, which have been found same as presented in \cite{d25}.

\section*{Acknowledgments}

``Authors thank the Higher Education Commission, Islamabad, Pakistan for its
financial support under the NRPU project with grant number
$\text{5329/Federal/NRPU/R\&D/HEC/2016}$''.

\end{document}